\documentclass{article}

\usepackage{times}
\usepackage{graphicx} % more modern
\usepackage{subfigure} 

\usepackage{natbib} % natbib requires that \bibliographystyle{} be set, else bibtex compile error exit 2

\usepackage{algorithm}
\usepackage{algorithmic}

\usepackage{hyperref}

\usepackage[accepted]{icml2017}

\icmltitlerunning{Learning non-linear motor control by FOLLOW}

\usepackage{amsmath,amssymb}
\usepackage{bm} % bold math

\begin{document}

\twocolumn[
\icmltitle{Non-linear motor control by local learning in spiking neural networks}



% you can specify symbols, otherwise they are numbered in order
\icmlsetsymbol{equal}{*}

\begin{icmlauthorlist}
\icmlauthor{Aditya Gilra}{to}
\icmlauthor{Wulfram Gerstner}{to}
\end{icmlauthorlist}

\icmlaffiliation{to}{School of Computer and Communication Sciences, and Brain-Mind Institute, School of Life Sciences, \'{E}cole Polytechnique F\'{e}d\'{e}rale de Lausanne, 1015 Lausanne EPFL, Switzerland}

\icmlcorrespondingauthor{Aditya Gilra}{aditya.gilra@epfl.ch}

\icmlkeywords{boring formatting information, machine learning, ICML}

\vskip 0.3in
]



\printAffiliationsAndNotice{}  % leave blank if no need to mention equal contribution

\begin{abstract}
Learning weights in a spiking neural network with hidden neurons, using local, stable and online rules, to control non-linear body dynamics is an open problem. Here, we employ a supervised scheme, Feedback-based Online Local Learning Of Weights (FOLLOW), to train a network of heterogeneous spiking neurons with hidden layers, to control a two-link arm so as to reproduce a desired state trajectory. The network first learns an inverse model of the non-linear dynamics, i.e. from state trajectory as input to the network, it learns to infer the continuous-time command that produced the trajectory. Connection weights are adjusted via a local plasticity rule that involves pre-synaptic firing and post-synaptic feedback of the error in the inferred command. We choose a network architecture, termed differential feedforward, that gives the lowest test error from different feedforward and recurrent architectures. The learned inverse model is then used to generate a continuous-time motor command to control the arm, given a desired trajectory.
\end{abstract}

\section{Introduction}

\paragraph{}
Motor control requires building internal models of the muscles-body system \citep{conant_every_1970,pouget_computational_2000,wolpert_computational_2000,lalazar_neural_2008}. The brain possibly uses random movements during pre-natal \citep{khazipov_early_2004} and post-natal development \citep{petersson_spontaneous_2003} termed motor babbling \citep{meltzoff_explaining_1997,petersson_spontaneous_2003} to learn internal models that associate body movements to neural motor commands 
\citep{lalazar_neural_2008,wong_can_2012,sarlegna_roles_2009,dadarlat_learning-based_2015}. For motor control, given a desired state trajectory for the dynamical system formed by the muscles and body, networks of spiking neurons in the brain must learn to produce the time-dependent neural control input that activates the muscles to produce the desired movement. Abstracting the muscles-body dynamics as
\begin{equation}
\label{eqn:dyn_sys}
d\vec{x}/dt=\vec{f}(\vec{x},\vec{u}),
\end{equation}
we require the network to generate control input $\vec{u}(t)$, given desired state trajectory $\vec{x}^D(t)$, that makes the state evolve as $\vec{x}(t) \approx \vec{x}^D(t)$. We train a network of spiking neurons with hidden layers, to learn the inverse model of the muscle-body dynamics, namely to infer control input $\vec{u}(t)$ given state trajectory $\vec{x}(t)$. The inverse model is then used to control the body, in closed loop mode.

\paragraph{}
In networks of continous-valued or spiking neurons, training the weights of hidden neurons with a local learning rule is considered difficult due to the credit assignment problem \citep{bengio_learning_1994,hochreiter_gradient_2001,abbott_building_2016}. Supervised learning in networks has typically been accomplished by backpropagation which is non-local \citep{rumelhart_learning_1986, williams_learning_1989}, reservoir computing which trains only output weights \citep{jaeger_echo_2001,maass_real-time_2002, legenstein_input_2003, maass_computational_2004,jaeger_harnessing_2004,joshi_movement_2005, legenstein_edge_2007}, FORCE learning which requires weight changes faster than the requisite dynamics \citep{sussillo_generating_2009,sussillo_transferring_2012,depasquale_using_2016,thalmeier_learning_2016,nicola_supervised_2016}, and other schemes for example \citep{sanner_gaussian_1992,dewolf_spiking_2016,macneil_fine-tuning_2011,bourdoukan_enforcing_2015}. Here, we employ a recent learning scheme called Feedback-based Online Local Learning Of Weights (FOLLOW) \citep{GilraPredictingnonlineardynamics2017}, which draws upon function and dynamics approximation theory \citep{funahashi_approximate_1989,hornik_multilayer_1989,girosi_networks_1990,sanner_gaussian_1992,funahashi_approximation_1993,pouget_spatial_1997,chow_modeling_2000,seung_stability_2000,eliasmith_neural_2004,eliasmith_unified_2005} and adaptive control theory \citep{morse_global_1980,narendra_stable_1980,slotine_adaptive_1986,slotine_adaptive_1987,narendra_stable_1989,sastry_adaptive_1989,ioannou_robust_2012}.
FOLLOW is a synaptically local and provably stable and convergent alternative for training the feedforward and recurrent weights of hidden spiking neurons in a network, that was employed to predict non-linear dynamics \citep{GilraPredictingnonlineardynamics2017}. While the authors of the FOLLOW scheme \citep{GilraPredictingnonlineardynamics2017} learned a forward model of arm dynamics, i.e. predicting arm state given neural control input, we train the network to perform full motor control, first learning the inverse model to infer control input given current arm trajectory, and then employing it in closed loop to make the arm replicate a desired trajectory.

\section{Network architecture and FOLLOW scheme to learn inverse model}

\begin{figure}
\includegraphics[width=\columnwidth]{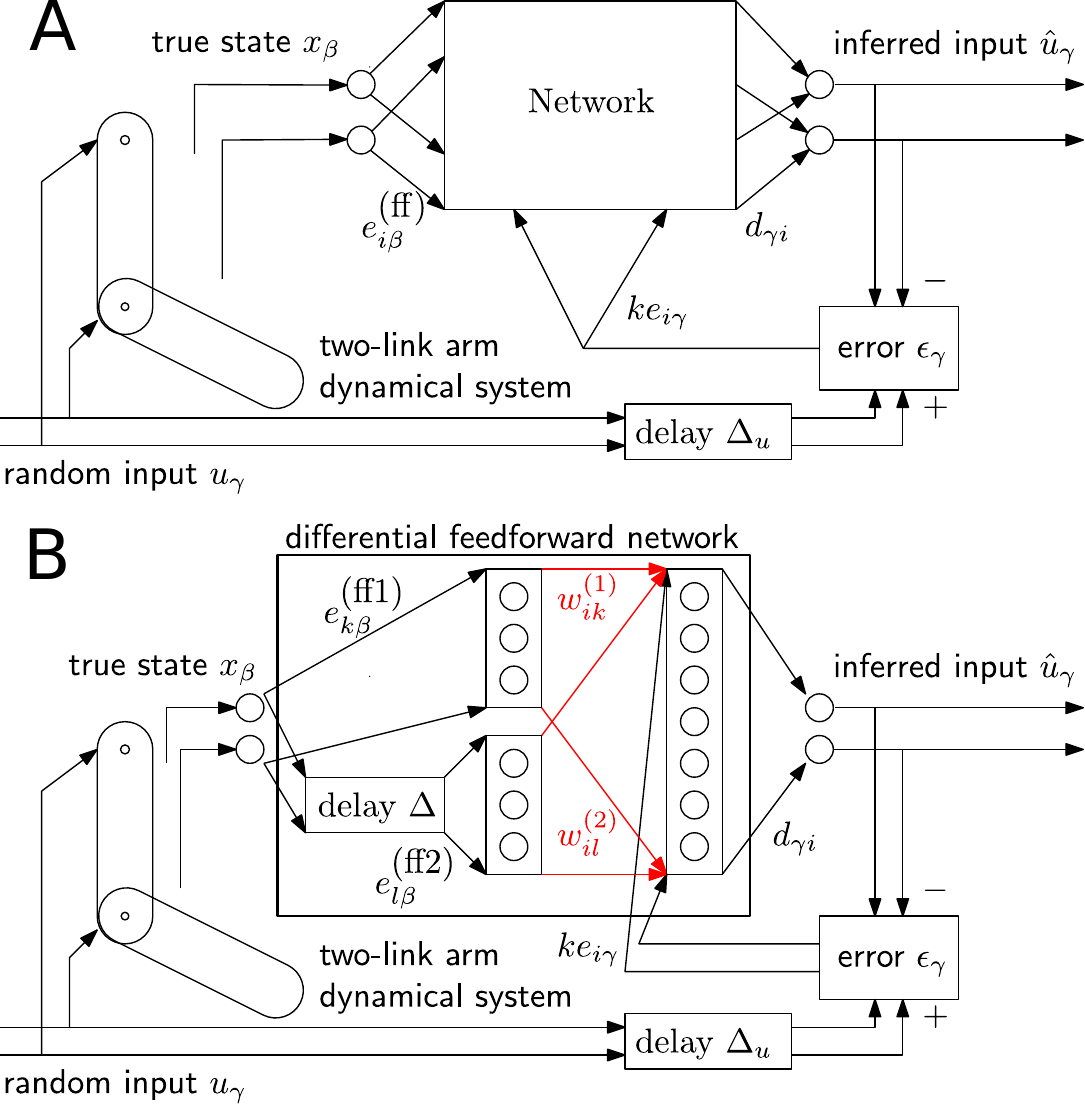}
\caption{\textbf{Network configuration for learning the inverse model.}
\textbf{A}. During learning, random motor commands cause arbitrary movements of the arm (motor babbling). Given the resulting state of the arm, the network must infer the continuous-time motor commands that caused it. The observed 4-dimensional state variables $x_\beta$ of the arm (which can be obtained from visual and proprioceptive feedback) are provided as input to the network via fixed random weights $e^{\textnormal{(ff)}}_{i\beta}$. The 2-dimensional motor command $\hat{u}_\gamma$ is linearly decoded from the filtered output spike trains of the network via fixed weights $d_{\gamma i}$. A copy of the random motor command (input to the arm) is used, after a delay of $\Delta_u$, to compute the error in the inferred motor command, i.e. error $\epsilon_\gamma(t) = u_\gamma(t-\Delta_u) - \hat{u}_\gamma(t)$. This deviation of the predicted command from the reference command, is fed back into the network with fixed random encoding weights $ke_{i\gamma}$. This error signal is used to update the network weights so as to infer the motor commands causing the observed state. Twin lines in the connection arrows denote multi-dimensional signals, but their number is not representative of the dimensionality. \textbf{B.} Learning inverse model with a differential feedforward network. The learning paradigm and configuration are as in panel A. The input is sent undelayed to a layer of neurons $\textnormal{ff1}$ via encoding weights $e^{\textnormal{(ff1)}}_{k\beta}$ and sent delayed by $\Delta$ to another layer of neurons $\textnormal{ff2}$ via encoding weights $e^{\textnormal{(ff2)}}_{l\beta}$. These two layers feed into the next `computation' layer with weights $w^{(1)}_{ik}$ and $w^{(2)}_{il}$ that are plastic (in red). The error is fedback to the `computation' layer with weights $ke_{i\gamma}$. This error projected into each neuron along with the filtered pre-synaptic spike train is used to update the feedforward weights in red.
\label{fig:schematic_inverse}
}
\end{figure}

\paragraph{}
We use a network of heterogeneous leaky-integrate-and-fire (LIF) neurons, with different biases to learn the inverse model of an arm, as depicted in Figure \ref{fig:schematic_inverse}. The arm, adapted from \citep{li_optimal_2006}, is modelled as a two-link pendulum moving in a vertical plane under gravity, with friction at the joints, $0^\circ$ as the equilibrium downwards position, and soft bounds on motion beyond $\pm 90^\circ$. As shown in Figure \ref{fig:schematic_inverse}A, random 2-dimensional torque input $u_\gamma(t), \gamma=1,2$ is provided to the arm to generate random state trajectories $\vec{x}(t)$ analogous to motor babbling. The 4-dimensional state of the arm (2 joint angles and 2 joint velocities), denoted $x_\beta, \beta=1,..,4$, is fed as input to the network via fixed, random weights. The network must learn to infer $\vec{u}(t)$ that generated the arm state trajectory $\vec{x}(t)$.

\paragraph{}
We chose the network in Figure \ref{fig:schematic_inverse}B, termed the differential feedforward network, from a variety of network architectures (see next section), to learn the inverse model. We have $K=L=3000$ neurons each in the two feedforward hidden layers indexed by $k=1\ldots K$, and $l=1\ldots L$, followed by $N=5000$ neurons in the `computation' hidden layer indexed by $i=1\ldots N$. The state vector $\vec{x}$ is fed to the two differential feedforward layers, undelayed to one and delayed to the other by interval $\Delta$, via fixed random weights $e_{k\beta}^{(\textnormal{ff1})}$ and $e_{l\beta}^{(\textnormal{ff2})}$ respectively. The current into a neuron $k$ in the undelayed hidden layer is given by 
\begin{equation}
J_k = \sum_\beta e^{\textnormal{ff1}}_{k\beta} x_\beta(t) + b_k,
\end{equation}
where $b_k$ is a fixed, random neuron-specific bias. Similarly, the current into a neuron $l$ in the delayed hidden layer is given by 
\begin{equation}
J_l = \sum_\beta e^{\textnormal{ff2}}_{l\beta} x_\beta(t-\Delta) + b_l,
\end{equation}
where $b_l$ is also a fixed, random neuron-specific bias.

\paragraph{}
The output $\vec{\hat{u}}(t)$ of the network is a linearly weighted sum of filtered spike trains $S_i(t)$ of the computation layer neurons:
\begin{equation}
\hat{u}_\gamma(t) = \sum_i d_{\gamma i} \int_{-\infty}^t S_i(s)\kappa(t-s)ds,
\end{equation}
with the filtering kernel $\kappa(t)$ a decaying exponential with time constant 20 ms, and the readout weights $d_{\gamma i}$. This readout $\vec{\hat{u}}(t)$ of the network is compared to a time-delayed version of the command input, with delay $\Delta_u$ ms, to causally infer the past command. Borrowing from adaptive control theory \citep{narendra_stable_1989,ioannou_robust_2012}, the error $\epsilon_\gamma \equiv u_\gamma(t-\Delta_u) - \hat{u}_\gamma(t)$ is fed back as neural currents to the computation layer neurons with fixed random feedback weights $e_{i\gamma}$ multiplied by a gain $k=10$. The total current into neuron $i$ in the `computation' layer, having fixed random bias $b_i$, is a sum of the feedforward1, feedforward2 and error currents:
\begin{multline}
\label{eqn:ff_adaptive_current}
J_i = \sum_k w^{(1)}_{ik} (S^\textnormal{ff1}_k*\kappa)(t) + 
        \sum_l w^{(2)}_{il} (S^\textnormal{ff2}_l*\kappa)(t) + \\
        \sum_\gamma k e_{i\gamma} (\epsilon_\gamma*\kappa)(t) + b_i.
\end{multline}

\paragraph{}
The trick in the FOLLOW scheme is to pre-learn the readout weights $d_{\gamma i}$ to be an auto-encoder with respect to error feedback weights $e_{i\gamma}$. Learning the auto-encoder can be accomplished by existing schemes \citep{voegtlin_temporal_2006,burbank_mirrored_2015,urbanczik_learning_2014}, and so the auto-encoder was pre-learned algorithmically here. Due to this auto-encoder, the error feedback with high gain $k$ acts as a negative feedback that serves to make the network output $\vec{\hat{u}}(t)$ \emph{follow} the true command torque $\vec{u}(t-\Delta)$, even without the hidden weights being learned. The system dynamics and command torque are required to vary slower than the synaptic timescale. As explained in \citep{GilraPredictingnonlineardynamics2017}, even the hidden neurons are entrained to fire as they ideally should, and this enables the feedforward (or recurrent) hidden weights to be learned, using only synaptically available quantities, as:
\begin{equation}
\frac{d w_{ij}}{dt} = \eta \,( I_i^\epsilon * \kappa^\epsilon ) (S_j*\kappa)(t),
\end{equation}
where $I^\epsilon_i \equiv k \sum_\alpha e_{i\alpha} \epsilon_\alpha$ is the error current injected into each neuron, $\kappa^\epsilon$ is a decaying exponential filter of time constant 200 ms and $S_j$ is the pre-synaptic spike train.

\paragraph{}
This FOLLOW learning rule is applied on the plastic weights (in red in Fig. \ref{fig:schematic_inverse}B), while the current state of the arm is fed to the network, with the network output following the time-delayed command input torque due to the error feedback loop. FOLLOW learning has been shown, under reasonable approximations, to be uniformly stable using a Lyapunov approach, with the squared error in the learned output tending asymptotically to zero \citep{GilraPredictingnonlineardynamics2017}. Details and parameters for generating the random input, the arm dynamics and the fixed random network parameters are as in \citep{GilraPredictingnonlineardynamics2017}.

\paragraph{}
After training the network on inferring the command input given the state trajectory during motor babbling, we expect that when we remove the error feedback loop, the network will still be able to infer the command input given the current trajectory. We show the various stages of learning and testing in Figure \ref{fig:inverse_diff-ff}, for the differential-feedforward network of Figure \ref{fig:schematic_inverse}B.

%\section{Learning two-link arm dynamics given input (forward model)}
%
%\begin{figure}
%\begin{center}
%\includegraphics[width=\textwidth]{motor_forward_schematic}
%\end{center}
%\caption{\textbf{Schematic for learning a forward model} \textbf{A.} 
%}
%\label{fig:schematic_forward}
%\end{figure}

\section{Learning inverse model: inferring control command given arm trajectory}

\begin{figure*}
\begin{center}
\includegraphics[width=0.8\textwidth]{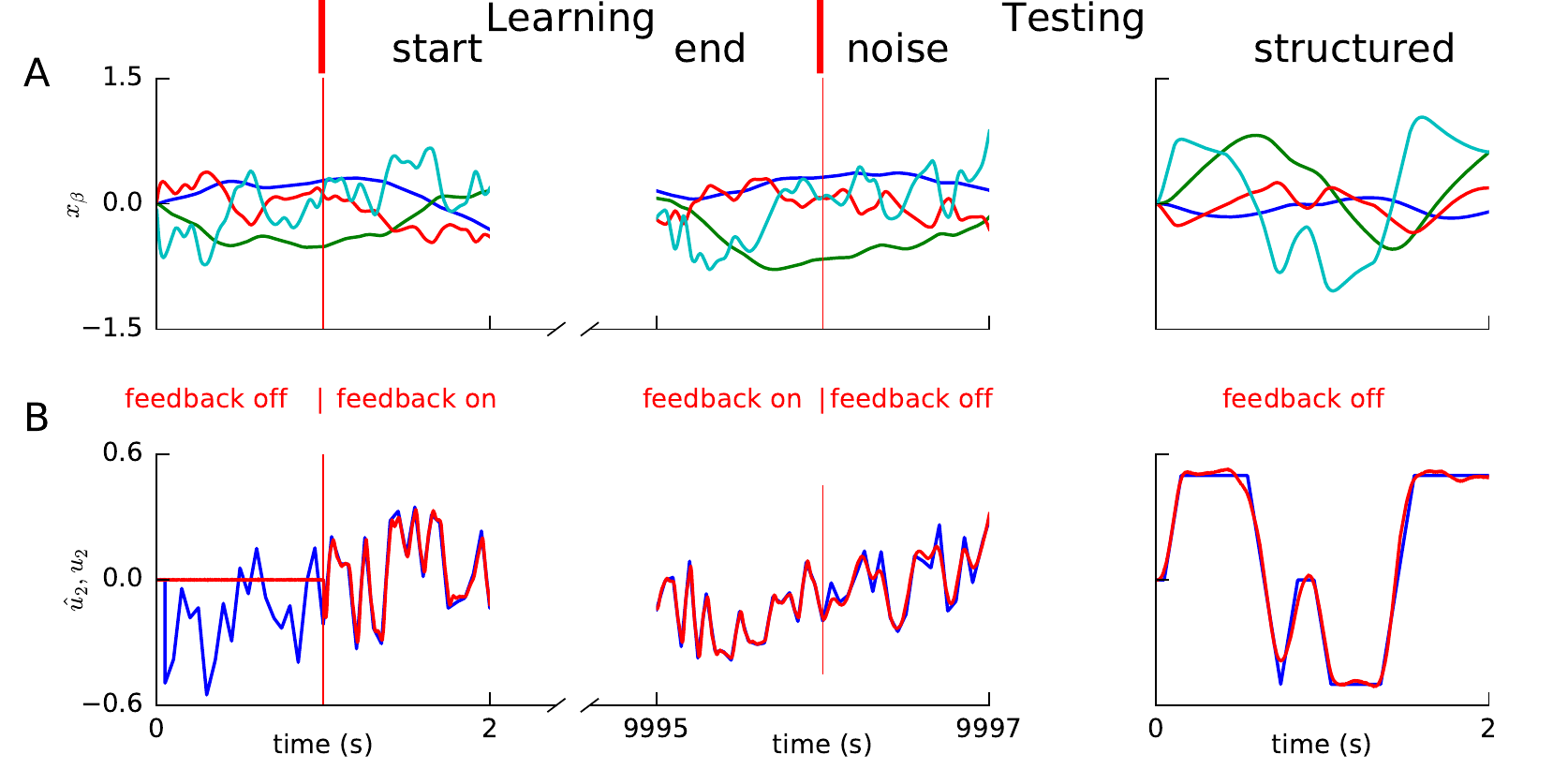}
\end{center}
\caption{\textbf{Stages during FOLLOW learning of inverse model using the differential feedforward network (Fig. \ref{fig:schematic_inverse}B).} \textbf{A,B}. The vertical red lines divide the figure into three time stages: before learning without feedback on the left, during learning with feedback in the middle, and during testing without feedback on the right. \textbf{A}. The input to the network $\vec{x}(t)$ namely the 4-dimensional state trajectories are shown in 4 different colours. \textbf{B}. A component of the reference torque $u_2(t-\Delta)$ to the arm (in blue) is shown along with the network readout $\hat{u_2}(t)$ (in red). Before learning without feedback (on the left), the network output (red) doesn't infer the command torque (blue). During learning with feedback (middle), the network output \emph{follows} the command torque due to the negative feedback (Fig. \ref{fig:schematic_inverse}), while the error, which decreases over time, is used to train the hidden weights. After learning (right), i.e. freezing weights and removing feedback, the network infers the command torque even without feedback, indicating that it has learned the inverse model.}
\label{fig:inverse_diff-ff}
\end{figure*}

\paragraph{}
At the start of learning, all trainable weights were initialized to zero. The state trajectory was fed as input to the network, while the network had to learn to produce the command (delayed by $\Delta_u$) that caused the trajectory.

\paragraph{}
In Figure \ref{fig:inverse_diff-ff}, we show the different stages in FOLLOW learning for the differential feedforward network (Fig. \ref{fig:schematic_inverse}B). Before learning, without feedback, the network output remained zero. With feedback on, even before the weights have been learned, the network output followed the desired output i.e. the time-delayed reference command. With feedback on, the network was trained with the FOLLOW learning rule. At the end of 10,000 s of FOLLOW learning on the feedforward weights, at learning rate 2e-3, we froze the weights and removed the feedback, in order to test if the network had learned the inverse model. We also tested the network on a more structured task trajectory, in addition to random motor babbling. The network output inferred the command given the state trajectory, even without feedback indicating that the network had learned the inverse model.

\begin{figure}
\begin{center}
\includegraphics[width=0.99\columnwidth]{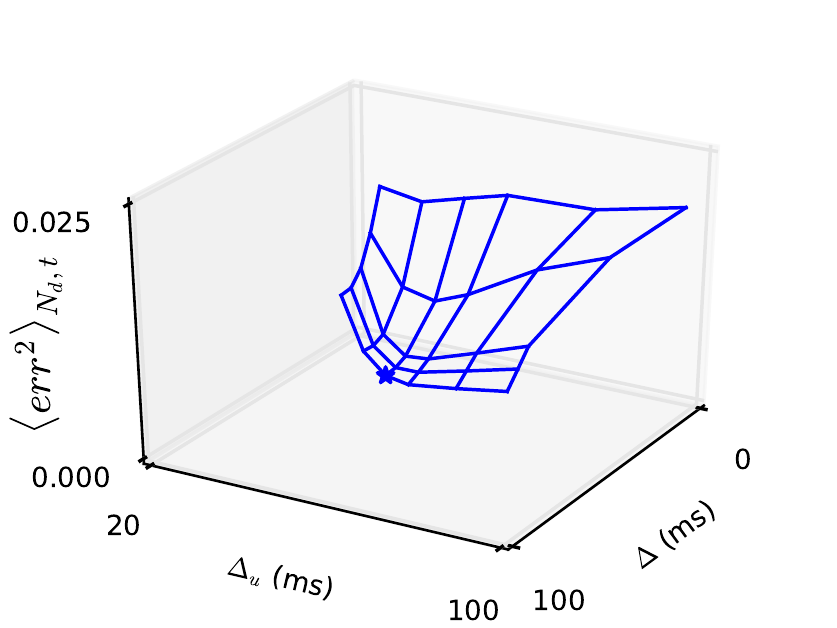}
\end{center}
\caption{\textbf{Mean squared test error versus $\Delta$ and $\Delta_u$.}
For different values of differential delay $\Delta$ and causal torque delay $\Delta_u$, we learned the inverse model using the differential feedforward network, having 200 neurons in each of the undelayed and delayed layers and 500 neurons in the computation layer. We plot the mean squared test error without feedback, after approximately 10,000s of learning, averaged over 4 s and per state dimension. The star marks the mean squared test error for $\Delta=50$ ms and $\Delta_u=50$ ms.
}
\label{fig:difftau_causaltau}
\end{figure}

\paragraph{}
We used a causal delay $\Delta_u$ in supplying the target commands, since the command in the past that caused the current state input needs to be inferred. The differential delay $\Delta$ between the undelayed and delayed feedforward layers possibly controls the accuracy of the temporal derivative. We swept the delays $\Delta_u$ and $\Delta$ over typical network time scales, shown in Figure \ref{fig:difftau_causaltau}, and found that $\Delta_u \approx 50$ ms and $\Delta \approx 50$ ms gave the lowest test error. This $\Delta_u$ is consistent with the delay due to two synaptic filtering time constants of $20$ ms each from the input state to the inferred command. For a command delayed by $\Delta_u = 50$ ms to be inferred, a $\Delta = 50$ ms differential delay allows the most accurate computation of the derivative, perhaps because it is closest to the time of the reference command, than a $\Delta = 10$ or $20$ ms delay, as seen in Figure \ref{fig:difftau_causaltau}. Thus, for all remaining simulations including Figure \ref{fig:inverse_diff-ff}, we used $\Delta=50$ ms and $\Delta_u=50$ ms.

\begin{figure}
\begin{center}
\includegraphics[width=\columnwidth]{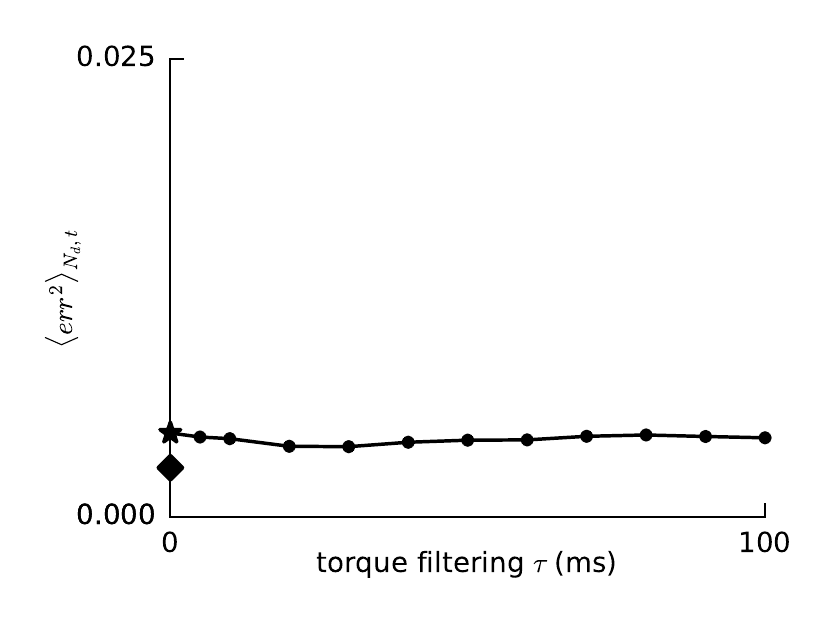}
\end{center}
\caption{\textbf{Mean squared test error versus torque filtering $\tau$.}
For increasing values of torque filtering time constant $\tau$, we learned the inverse model using the same differential feedforward network as in Figure \ref{fig:difftau_causaltau}, with $\Delta=50$ ms and $\Delta_u=50$ ms. We plot the mean squared test error without feedback, as in Figure \ref{fig:difftau_causaltau}, with the same range for comparison. The star for $\tau=0$ marks the mean squared error for the same simulation as the star in Figure \ref{fig:difftau_causaltau}, while the diamond below it marks the mean squared error for the simulation with larger number of neurons in Figure \ref{fig:inverse_diff-ff}.
}
\label{fig:torquefilt}
\end{figure}

\paragraph{}
We asked whether the seemingly low-pass filtering of the inferred torque during testing in Figure \ref{fig:inverse_diff-ff}B, was because the command varied faster than what the network could approximate. We used low-pass filtered commands for both the arm and the network, with increasing filtering time constants, using a decaying exponential kernel, but the test error remained almost the same, as shown in Figure \ref{fig:torquefilt}. With a larger number of neurons, the test error came down without any filtering on the torque (Fig. \ref{fig:torquefilt}), so the approximation error due to finite number of neurons seemed larger than that due to any fast timescales in the torque. Thus, in all further simulations, we did not use any filtering on the torque. %Thus we conjecture that while the network is able to compute the different $\Delta \vec{x}$, it is unable to invert the function $f$ as there isn't another hidden layer to compute the inversion. A hierarchical FOLLOW learning scheme to overcome this lacuna is an interesting direction to pursue.

\paragraph{}
The network approximates the inverse dynamics using the tuning curves of the heterogeneous neurons as an overcomplete set of basis functions \citep{funahashi_approximate_1989,hornik_multilayer_1989,girosi_networks_1990,sanner_gaussian_1992,funahashi_approximation_1993,pouget_spatial_1997,chow_modeling_2000,seung_stability_2000,eliasmith_neural_2004,eliasmith_unified_2005}. As studied in adaptive control theory \citep{ioannou_robust_2012,narendra_stable_1989,ioannou_adaptive_2006}, the error due to approximation causes a drift in weights, which can lead to error increasing after some training time. The same literature also suggests ameliorative techniques which include a weight leakage term switched on slowly, whenever a weight crosses a set value \citep{ioannou_robust_1986,narendra_stable_1989}, or a dead zone policy of not updating weights if the error falls below a threshold \citep{slotine_adaptive_1986,ioannou_robust_2012}. However, we did not need to implement these techniques for learning the inverse model with our networks, suggesting that the approximations with our finite number of neurons were reasonable.

\begin{figure}[ht]
\includegraphics[width=\columnwidth]{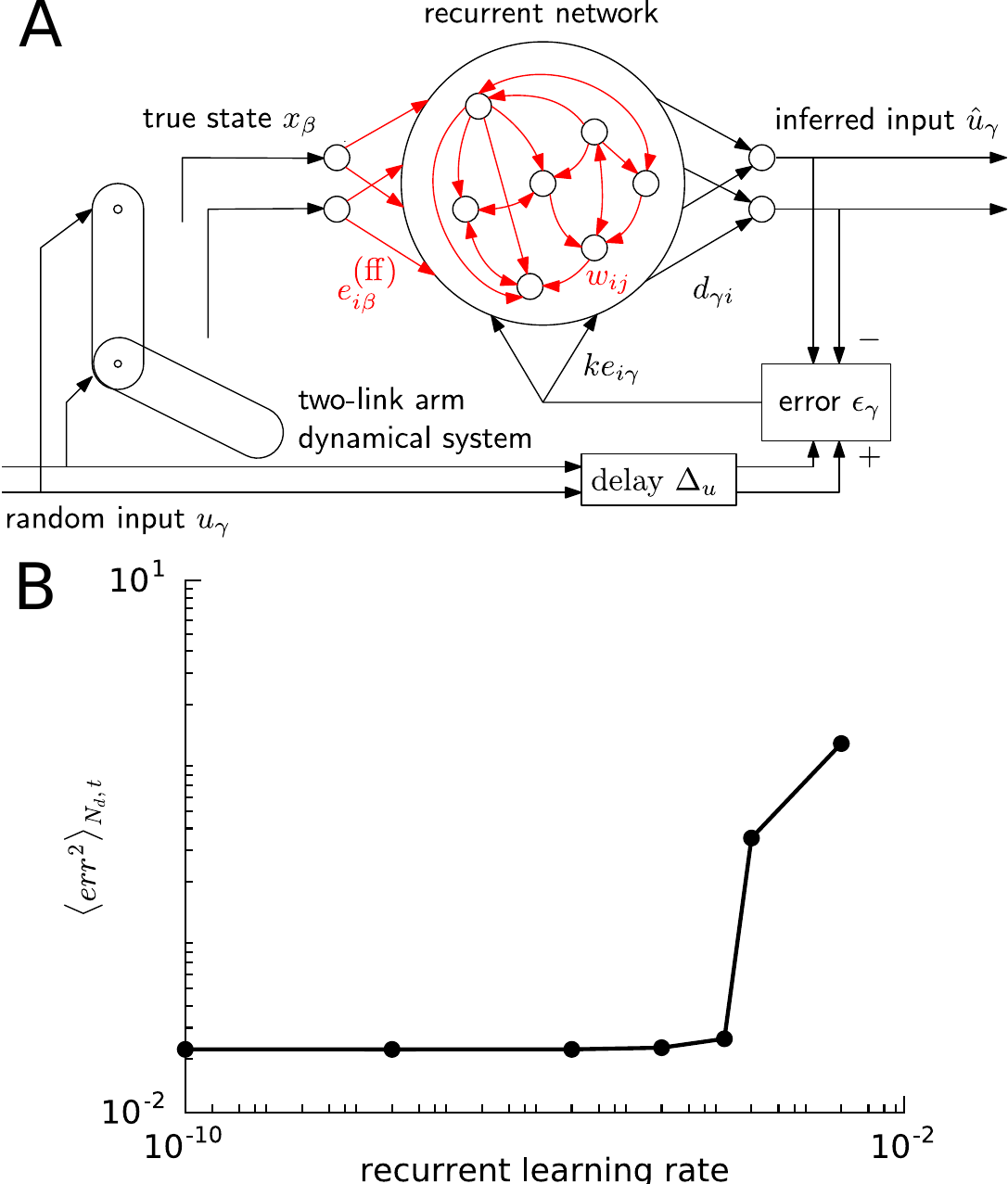}
\caption{\textbf{Recurrent network cannot learn the inverse model via FOLLOW learning.}
\textbf{A}. The learning configuration is same as Figure \ref{fig:schematic_inverse}A, except that the spiking network is recurrently connected. The feedforward weights $e_{i\beta}^{\textnormal{ff}}$ and the recurrent weights $w_{ij}$ are plastic under the FOLLOW rule. \textbf{B}. We trained the recurrent network on the inverse model as in A, with 900 neurons in the recurrent layer and $\Delta_u=50$ ms, at increasing values of the learning rate (x-axis, log scale) on the recurrent weights, while the feedforward weights were learned at a constant rate. We plot the mean squared test error (y-axis, log scale) without feedback, after approximately 500s of learning, averaged over 4 s and per state dimension.
}
\label{fig:reclearning}
\end{figure}

\paragraph{}
In learning the forward model, the command $\vec{u}$ must be integrated as $\int f(\vec{x},\vec{u}) dt$ to obtain the state $\vec{x}$, see equation \eqref{eqn:dyn_sys}. However, for the inverse model, the state $\vec{x}$ has to be differentiated, and the function $f$ inverted, to infer the command $\vec{u}$. In principle, we could use a recurrent network, as it is Turing complete \cite{SiegelmannComputationalPowerNeural1995}, to approximate the inverse model, provided all the input, recurrent and output weights are wired correctly. However, in the FOLLOW learning scheme \citep{GilraPredictingnonlineardynamics2017}, the output weights are fixed to match the error encoding weights (auto-encoder), to enable local learning. Indeed, we tried to learn the inverse model with a recurrent network as in Figure \ref{fig:reclearning}A. The total current into neuron $i$ in the recurrent layer, having fixed random bias $b_i$, was a sum of the feedforward, recurrent and error currents:
\begin{multline}
\label{eqn:adaptive_current}
J_i = \sum_\beta e^{\textnormal{ff}}_{i\beta} x_\beta(t) +
        \sum_j w_{ij} (S_j*\kappa)(t) + \\
        \sum_\gamma k e_{i\gamma} (\epsilon_\gamma*\kappa)(t) + b_i.
\end{multline}
We kept a fixed learning rate on the feedforward weights, but increased the learning rate on the recurrent weights from effectively no learning to a reasonable learning rate 2e-3 that had worked very well for learning the forward model using a recurrent network \citep{GilraPredictingnonlineardynamics2017}. As shown in Figure \ref{fig:reclearning}, we found that test performance dropped sharply with learning on the recurrent weights. Near zero learning on the recurrent weights gave the lowest mean squared test error. Even if we included a feedforward layer between the state input and the recurrent network, with learnable feedforward and recurrent weights, as for learning the forward model in \citep{GilraPredictingnonlineardynamics2017}, still test performance degraded with learning on the recurrent weights. Note that a recurrent network in the FOLLOW learning scheme, which has fixed output weights (Fig. \ref{fig:reclearning}A), cannot be transformed into the differential feedforward architecture (Fig. \ref{fig:schematic_inverse}B). Even if part of the recurrent network can learn to delay or hold an earlier state in memory, it always remains connected to the output by fixed weights.

\begin{figure*}
\begin{center}
\includegraphics[width=\textwidth]{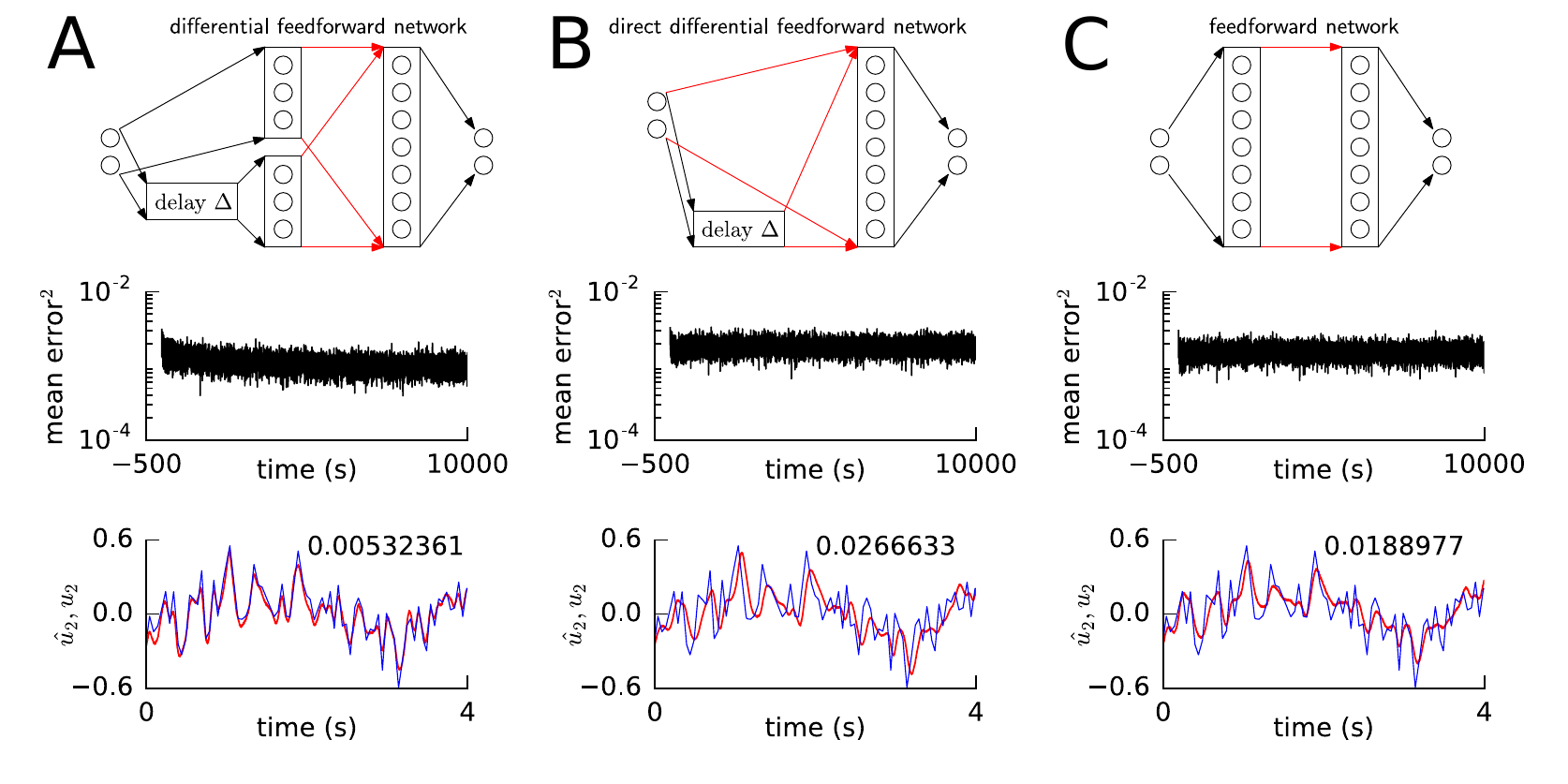}
\end{center}
\caption{\textbf{Comparison of feedforward network architectures.}
\textbf{A,B,C}. In the top panel, the feedforward network architecture used to learn the inverse model is shown. Weights in red are plastic. Each network has the same number of heterogeneous LIF neurons. Mean squared error during training with feedback is plotted versus time over approximately 10,000s seconds of learning in the middle panels. Reference torque component $u_2$ (blue) and inferred torque component $\hat{u}_2$ (red) given desired state are plotted versus time, during 4 s of testing without feedback. The mean squared test error without feedback per unit time per state dimension is written above the plot. Mean squared error during test without feedback is typically larger than that during training due to corrective negative feedback. \textbf{A}. The differential feedforward network is our default network architecture, with 200 neurons in the undelayed and delayed layers and 500 neurons in the computation layer, with $\Delta=50$ ms and $\Delta_u=50$ ms. \textbf{B}. The direct differential feedforward network doesn't have intermediate undelayed and delayed hidden layers. Rather the undelayed and delayed states are directly feed on to the computation layer having 900 neurons. \textbf{C}. The purely feedforward network has 450 neurons in each hidden layer.
}
\label{fig:archi_compare}
\end{figure*}

\begin{figure*}
\begin{center}
\includegraphics[width=0.8\textwidth]{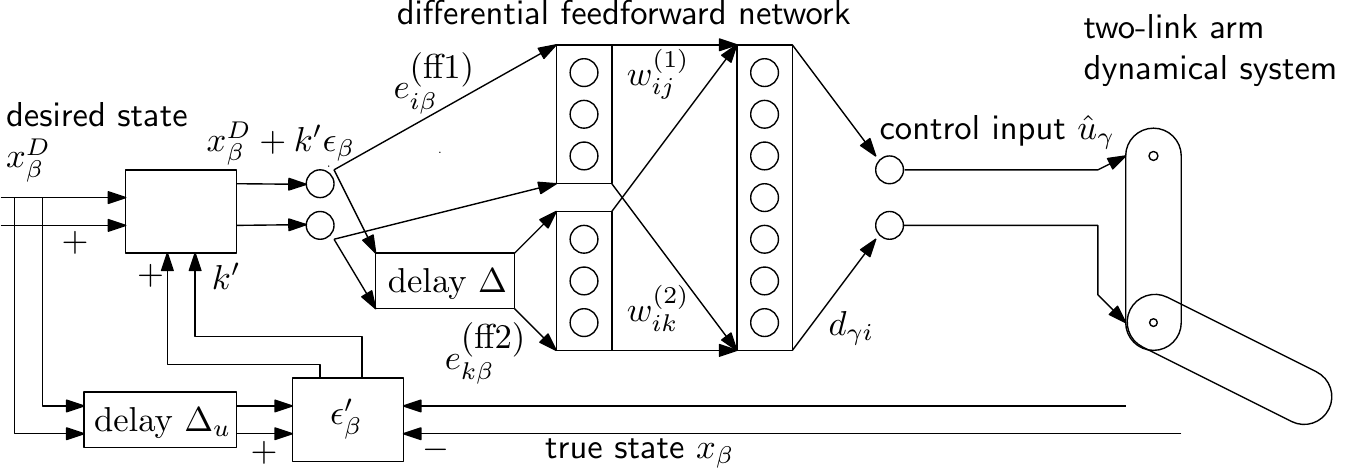}
\end{center}
\caption{\textbf{Schematic for control using inverse model.}
The desired state trajectory $\vec{x}^D(t)$ is fed into the differential feedforward network that is already trained as an inverse model. The control input $\hat{u}_\gamma$ is read out from the network. The read out control input torques are delivered to the arm which produces the true state $x_\beta(t)$, which is already close to the desired one if the inverse model performs well. To further improve long-time performance, the true state of the arm is compared with the desired state, delayed by $\Delta_u = 50$ ms (since the inverse model has learned to infer the control input with this delay), and the error $\epsilon'_\beta = (x^D_\beta - x_\beta)$ is fed back with gain $k'=3$. The non-linear feedforward control by the inverse model brings the arm state close to the desired state, after which the linear negative feedback control brings the arm state even closer.  Twin lines in the connection arrows denote multi-dimensional signals, but their number is not representative of the dimensionality.
}
\label{fig:schematic_control}
\end{figure*}

\paragraph{}
Further, even from among the feedforward network architectures we tried with FOLLOW learning, as shown in Figure \ref{fig:archi_compare}, we found that the best performace for a fixed number of neurons was obtained by the differential feedward network (Fig. \ref{fig:schematic_inverse}B). Compared to the other feedforward architectures, we imagine that the differential feedforward network could learn to compute a temporal difference between the undelayed and delayed states, approximating the time derivative.

\section{Moving the arm as desired using the inverse model}

\paragraph{}
Having learned the inverse model, we used it to control the arm to reproduce a desired state trajectory, for which the control input was not known and had to be generated by the network. As shown in Figure \ref{fig:schematic_control}, the desired state $x^D_\beta(t)$ was fed to the network which had already been trained as an inverse model to output the requisite control command $\hat{u}_\gamma$. This command generated by the network was fed as torques to the joints of the arm to produce the desired motion. In this simple feed-forward or open loop mode, small errors in the generated control command will integrate over time into larger errors in the trajectory of the arm, causing large deviations from the desired trajectory over time. To ameliorate this, we closed the loop and injected the difference in the desired state and the true state multiplied by gain $k'=3$, back into the network input (Fig. \ref{fig:schematic_control}). In computing the error, the desired state was delayed by $\Delta_u=50$ ms, since the network was also trained to output the control command with this lag. This feedback loop for the error in the state variables, fed into the network via the input weights during control, is different from the feedback loop in the error for control command, fed back via the error feedback weights during learning. The latter feedback loop is in fact used only during learning and is not present during control.

\begin{figure}[t]
\begin{center}
\includegraphics[width=0.9\columnwidth]{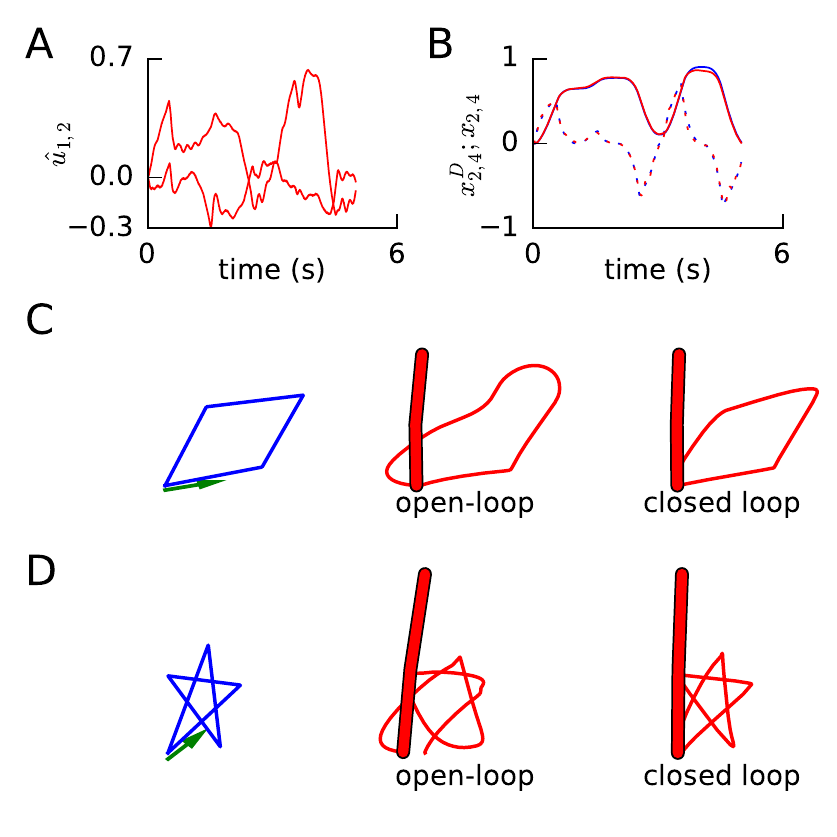}
\end{center}
\caption{\textbf{Control of two-link arm via inverse model.}
\textbf{A.} The elbow and shoulder control torques generated by the learned differential feedforward network under closed loop in the control architecture (Fig. \ref{fig:schematic_control}) in reponse to a desired star-drawing trajectory. \textbf{B.} The star-drawing trajectories for the elbow angle (solid) and angular velocity (dashed) as desired (blue) and as produced by the arm (red) in response to the torques in \textbf{A}, under closed loop control of the inverse model. \textbf{C,D.} The network controls the arm to draw a diamond (C) and a star (D). On the left in blue, is the desired trajectory, in coordinate space, that the end-point of the arm is required to trace, starting in the direction of the green arrow. In the middle in red, is the trace produced by the arm under open-loop control, i.e. no feedback of the error in state back to the network. To the right, in red is the trace produced under closed-loop, with feedback (gain $k'=3$) of the error in state variables.
}
\label{fig:control_test}
\end{figure}

\paragraph{}
We tested our motor control scheme for drawing a diamond and a star on the wall (Fig. \ref{fig:control_test}). In open loop, i.e. not feeding back the error in state variables, the command torque generated by the network enabled the arm to draw a pattern similar to that desired (Fig. \ref{fig:control_test} C,D and Supplementary movies 1 and 2). However, some parts of the pattern, which required large bending of the elbow joint were difficult to reproduce, possibly due to poorer learning at the limits of the motion, and divergence of the state from the desired state in open loop over time.

\paragraph{}
When we closed the loop with gain $k'=3$ the pattern was reproduced better (Fig. \ref{fig:control_test} C,D and Supplementary movies 3 and 4). Here, we used only proportional feedback with low gain. A proportional-integral-derivative (PID) feedback should provide even better tracking of the output without oscillations and offset.

\paragraph{}
Without the inverse model, the linear negative feedback loop (be it proportional or PID) on the state variables cannot make the arm follow the desired trajectory, even by tuning feedback gains, because of the complicated non-linear dynamics of the two-link arm. In our control scheme, the intuition is that the inverse model learned in the weights of the network produces an open-loop command that brings the arm close enough to the desired state, such that linear negative feedback around this momentary operating point can bring the arm even further closer.

\section{Discussion}

\paragraph{}
We used a synaptically local, online and stable learning scheme, FOLLOW \citep{GilraPredictingnonlineardynamics2017}, to train a network of heterogeneous spiking neurons with hidden layers, to infer the continuous-time command needed to drive a non-linear dynamical system, here a two-link arm, to produce a desired trajectory. We found that the differential feedforward network architecture performed best in learning the inverse model, from among a variety of feedforward and recurrent architectures, under the FOLLOW learning scheme. Under closed-loop, this trained inverse model was then used to control the arm. We expect that with proportional-integral-derivative feedback, the control of the arm will improve further.

\paragraph{}
Previous methods of training networks with hidden units incorporate only some of the desirable features of FOLLOW learning. Backpropagation and variants, backpropagation through time (BPTT) \citep{rumelhart_learning_1986} and real-time recurrent learning (RTRL) \citep{williams_learning_1989}, are non-local in time or space \citep{pearlmutter_gradient_1995,jaeger_tutorial_2005}. Synaptically local learning rules in recurrent spiking neural networks have typically not been demonstrated for learning inverse models and non-linear motor control \citep{macneil_fine-tuning_2011,bourdoukan_enforcing_2015,GilraPredictingnonlineardynamics2017,AlemiLearningarbitrarydynamics2017}. Reservoir computing methods \citep{jaeger_echo_2001,maass_real-time_2002,legenstein_input_2003,maass_computational_2004,jaeger_harnessing_2004,joshi_movement_2005,legenstein_edge_2007} did not initially learn weights within the hidden units, while newer FORCE learning methods that did \citep{sussillo_generating_2009, sussillo_transferring_2012,depasquale_using_2016,thalmeier_learning_2016,nicola_supervised_2016}, require weights to change faster than the time scale of the dynamics, and require different components of a multi-dimensional error to be fed to different sub-parts of the network. Reward-modulated Hebbian rules have not yet been used for continuous-time control \citep{legenstein_reward-modulated_2010,hoerzer_emergence_2014,kappel_reward-based_2017}. Approaches to control of non-linear systems seem to use non-local rules \citep{sanner_gaussian_1992,slotine_adaptive_1987,slotine_adaptive_1986,dewolf_spiking_2016,zerkaoui_stable_2009,hennequin_optimal_2014,song_training_2016} or use abstract networks \citep{berniker_deep_2015,hanuschkin_hebbian_2013}.

\paragraph{}
Since FOLLOW-based learning of motor control uses spiking neurons and is local and online, it is biologically plausible in these aspects, as a candidate learning scheme for motor control in the brain. However, it violates Dale's law in that outgoing synaptic weights from a single neuron in the learned network can be both positive and negative. However, even without obeying Dale's law, it can be implemented directly in neuromorphic hardware and incorporated into neuro-robotics for motor control, where the spike-based coding provides power efficiency, and localility eases hardware implementation and speeds up computations required for learning.

\paragraph{}
Interesting directions to explore could be to learn and control more complex dynamical systems like robotic arms in real-time, to implement the same on neuromorphic hardware, to make the FOLLOW scheme more biologically plausible, to learn to generate the desired trajectory given a higher-level goal via reinforcement learning, and to extend motor control to a hierarchy of levels as in the brain.

\section{Acknowledgements}
\paragraph{}
Financial support was provided by the European Research Council (Multirules, grant agreement no. 268689), the Swiss National Science Foundation (Sinergia, grant agreement no. CRSII2\_147636), and the European Commission Horizon 2020 Framework Program (H2020) (Human Brain Project, grant agreement no. 720270).

\bibliographystyle{icml2017}
\bibliography{zotero_neuroscience}
%\bibliography{/home/aditya/LIN_DATA/2015_Wulfram_lab/2015_pop_rand_coding/paper/zotero_neuroscience.bib}

\end{document}